\documentclass[conference]{IEEEtran}
\usepackage{graphicx,authblk,hyperref,float,caption,subcaption,url,booktabs,tabularx,array}
\newcolumntype{L}{>{\raggedright\arraybackslash}X}
\usepackage[backend=biber,style=numeric]{biblatex}
\title{Head Coil-Mounted Vision Correction Device for Magnetic Resonance Imaging}
\author[1,2]{Aleksandar Marinkovic}
\author[2,3]{Julian Michael Tyszka}
\affil[1]{Division of Engineering and Applied Science, California Institute of Technology, Pasadena, California}
\affil[2]{Division of Humanities and Social Science, California Institute of Technology, Pasadena, California}
\affil[3]{Caltech Brain Imaging Center, California Institute of Technology, Pasadena, California}

\begin{document}

\maketitle

\section{Abstract}
Correction for nearsightedness and farsightedness is an important concern for functional magnetic resonance imaging (fMRI) experiments involving visual stimuli in humans. In the absence of personal contact lenses, spherical refractive errors are typically corrected using interchangeable lenses mounted in goggles or glasses frames worn by the participant, or mounted on the head coil during scanning. The coil-mounted device described here avoids the ergonomic challenges encountered with head-mounted goggles and addresses limitations of prior coil-mounted designs, including ease of lens switching and inter-pupillary distance adjustments. Our device can be 3D printed economically with MRI-compatible plastics, including PLA.
\\\\
\textbf{Keywords:} vision correction, magnetic resonance imaging, MRI, fMRI, 3D printing, computer-aided design (CAD), computer-aided manufacturing (CAM)
\\\\
\texttt{Corresponding Author: Aleksandar Marinkovic (aleks@caltech.edu)}

\begin{table}[H]
\small
\renewcommand{\arraystretch}{1.2}
\centering
\begin{tabularx}{\columnwidth}{@{}
  >{\raggedright\arraybackslash}p{0.31\columnwidth}
  >{\raggedright\arraybackslash}p{0.60\columnwidth}
  X @{}}
\toprule
\textbf{Specifications Table}  \\
\midrule
Hardware Name & Head Coil-Mounted Vision Correction Device for Magnetic Resonance Imaging \\ 
Hardware Type & Optical device \\
Open Source License & CC BY-NC-SA 4.0 \\
Cost of hardware & \$25 \\
Source file repository & \href{https://github.com/aleks123marinkovic/VisionCorrectionDevice}{github link} \\
\bottomrule
\end{tabularx}
\normalsize
\end{table}

\section{Introduction} \label{intro}
Many adults living in the United States require some form of vision correction for refractive errors. The 1999-2004 National Health and Nutrition Examination Survey (NHANES) estimates the proportions of US adults with long-sightedness (hyperopia), short-sightedness (myopia) and astygmatism (cylindrical errors) at approximately 4\%, 33\% and 36\% respectively \cite{Vitale2008-hc}. As a consequence, some form of vision correction is generally required for functional MRI experiments involving visual stimuli in adult volunteers and generally involves adding MRI-compatible corrective lenses in the line of sight between the stimulus display and the participant. Typical correction systems involve either a head- or coil-mounted lens holder supporting a range of interchangeable lenses correcting for spherical errors between approximately -5 and +5 diopters in half-diopter increments. To our knowledge, no commercial, MRI-compatible device exists for correction of cylindrical (astigmatic) or higher order errors. While personal prescription glasses are not guaranteed to be MRI-compatible due to metallic frame and fastener materials, personal prescription contact lenses are a favored alternative to head or coil mounted interchangeable lens systems, and offer more accurate correction while still being compatible with fMRI visual tasks. However not all participants have access to or can tolerate contact lenses.
\\\\
Existing head-mounted vision correction devices most commonly use flexible plastic goggles with an elastic strap securing the goggles to the participant's head. The strap may become uncomfortable during longer scans and the goggles may contact the anterior element of the head coil in participants with larger head sizes, which in turn exerts uncomfortable pressure on the participant's face \cite{practiCal_fMRI_2016}.
\\\\
Coil-mounted vision correction devices address goggle discomfort issues by mounting corrective lenses on the coil hardware with no contact with the participants head or face. However, existing coil-mounted devices do not allow for variation in inter-pupillary distance (IPD) which in turn can reduce the accuracy of the vision correction with general-purpose spherical correction lenses.
\\\\
We describe a 3D-printable lens holder for vision correction during magnetic resonance imaging of the head. The device mounts securely on the anterior or upper element of a commercial head coil and allows for rapid exchange of lenses while accommodating individual inter-pupillary distances. The specific design described here was developed specifically for a widely-used commercial head coil (32 channel head array, Siemens Medical Solutions) but can be readily adapted to the geometry of head coils from other manufacturers, including General Electric and Philips.

\subsection{Prior Art}
Previous examples of commercial and non-commercial head and coil-mounted vision correction systems are summarized below.

\subsubsection{practiCal fMRI Coil-Mounted Lens Holder} 
A February 2016 entry in the practiCal fMRI online blog \cite{practiCal_fMRI_2016} describes a 3D printable vision corrector consisting of two 3D printed lens mounts that drop into the eye openings of the anterior element of the head coil with an optional securing strip. Fabrication details are provided in a later October 2018 entry in the techniCal fMRI blog \cite{techniCal_fMRI_2016}. This device allows for convenient lens switching but does not support IPD adjustment since lens position in the mounts and mount position in the eye openings are both fixed. The device appears to be generalizable to different commercial lens sets, such as those commonly included in head-mounted goggle systems.

\subsubsection{Cambridge Research Systems Coil-Mounted Device}
Cambridge Research Systems Ltd. (Rochester, Kent, UK) market two vision correction devices: the MRIFocus\textregistered\ vision correction device for the Siemens 3T head coil and the Nova Medical Lens Mount for the Nova Medical 7T head coil. The MRIFocus device mounts on the patient mirror gantry which in turn mounts to the anterior element of the head coil \cite{CRS_MRIFocus}. The Nova Medical device mounts inside the 7T head coil behind the nose piece. Neither device allows for IPD adjustment and both devices require the use of an included custom corrective lens set.

\subsubsection{Head Mounted Goggle Systems} \label{goggles}
At time of writing, and to the best of our knowledge, two companies market head-mounted goggle-based vision correction systems for fMRI. Cortech Solutions (Wilmington, NC) market a range of MediGlasses\textregistered\ vision correction kits \cite{Cortech_Mediglasses} consisting of goggles and interchangeable lenses. A similar correction system is marketed by Psychology Software Tools (Pittsburgh, PA) consisting of a single google and lens set \cite{PST_Lenses}. Both systems have ergonomic issues for longer scans such as those commonly used in research, and for smaller, tight-fitting head coils (see Introduction).

\subsection{Vision Correction Device with Inter-Pupillary Distance Adjustment}
We have developed a new vision correction device for fMRI which addresses the lack of IPD adjustment in existing coil-mounted devices, is fully customizable for different commercial lens set geometries and avoids the ergonomic issues of head-mounted goggle systems in small, tight head coils. IPD adjustment in the new device is achieved through independent horizontal motion of the lens holders, limited only the geometry of the lenses and width of the head coil mount slots.

\section{Mechanical Design}
\begin{figure}[H]
  \centering
  \includegraphics[width=0.5\textwidth]{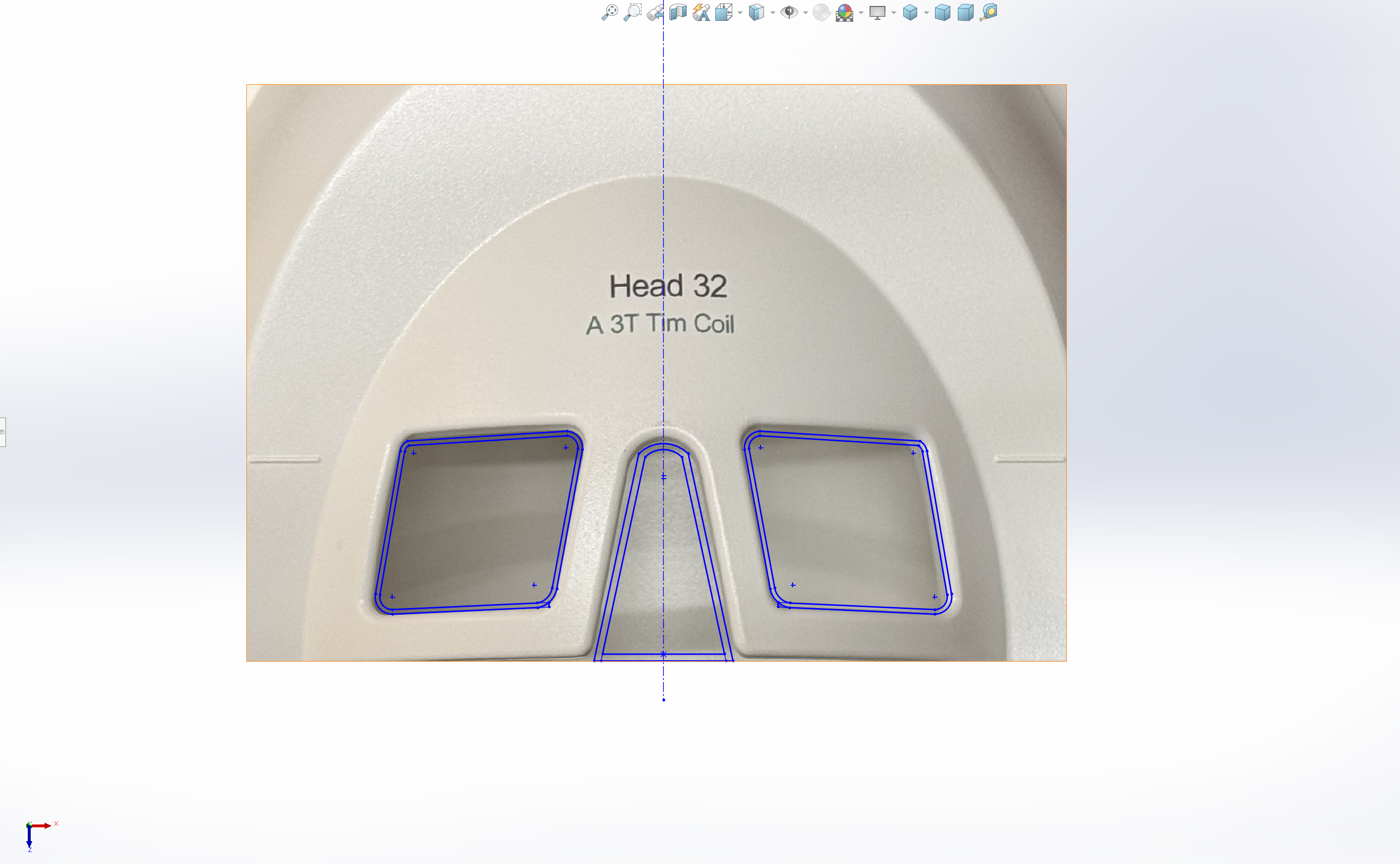}
  \caption{Example of the use of the sketch picture tool in SolidWorks to obtain the geometry of the anterior Head Coil and design the Head Coil Mount.}
  \label{fig:Sketch}
\end{figure}

\begin{figure}[H]
  \centering
  \begin{subfigure}[b]{0.4\textwidth}
    \centering
    \includegraphics[width=\textwidth]{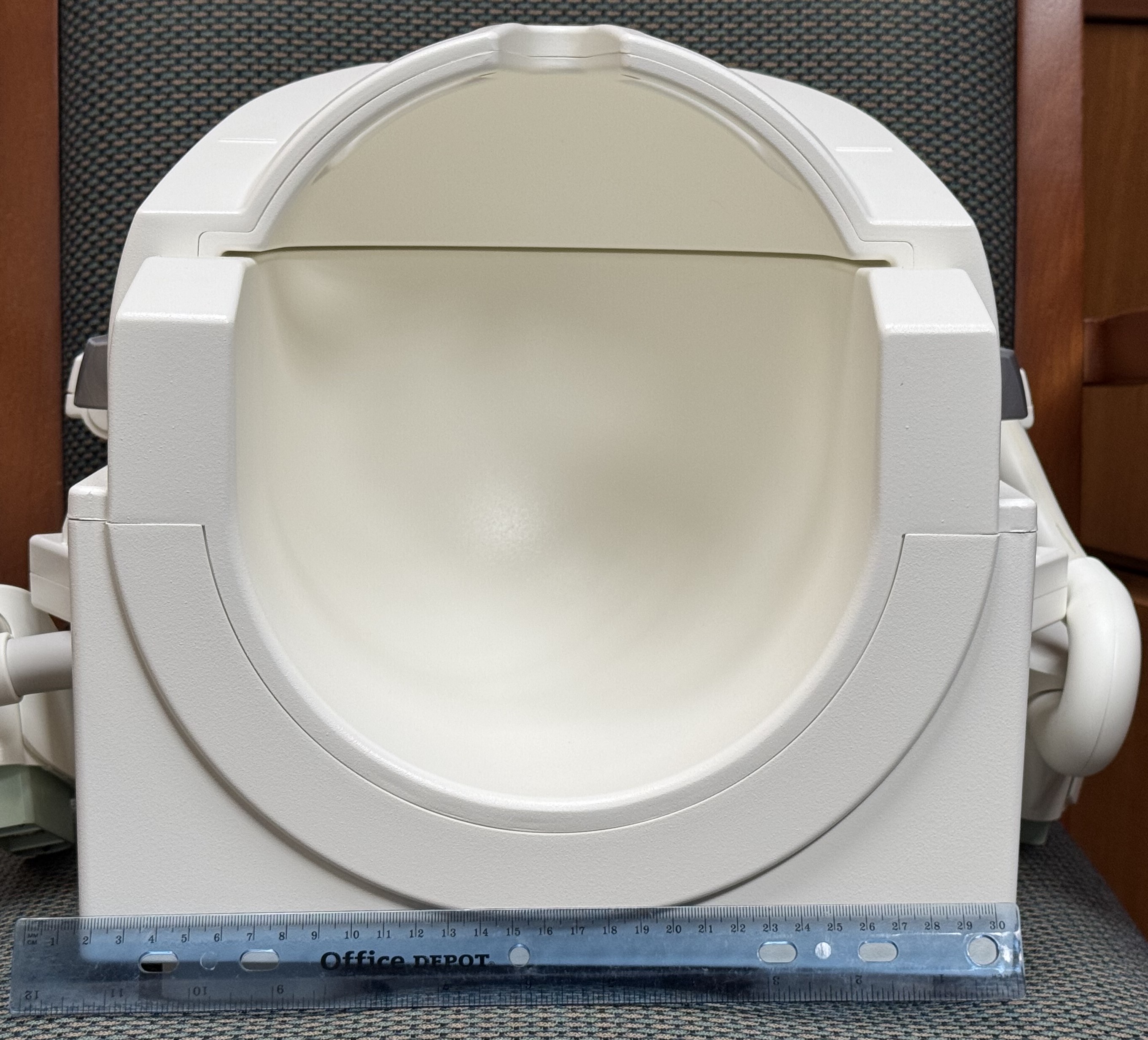}
    \caption{Front view of Head Coil}
    \label{fig:sub_headcoilmount_drawing}
  \end{subfigure}
  \hfill
  \begin{subfigure}[b]{0.4\textwidth}
    \centering
    \includegraphics[width=\textwidth]{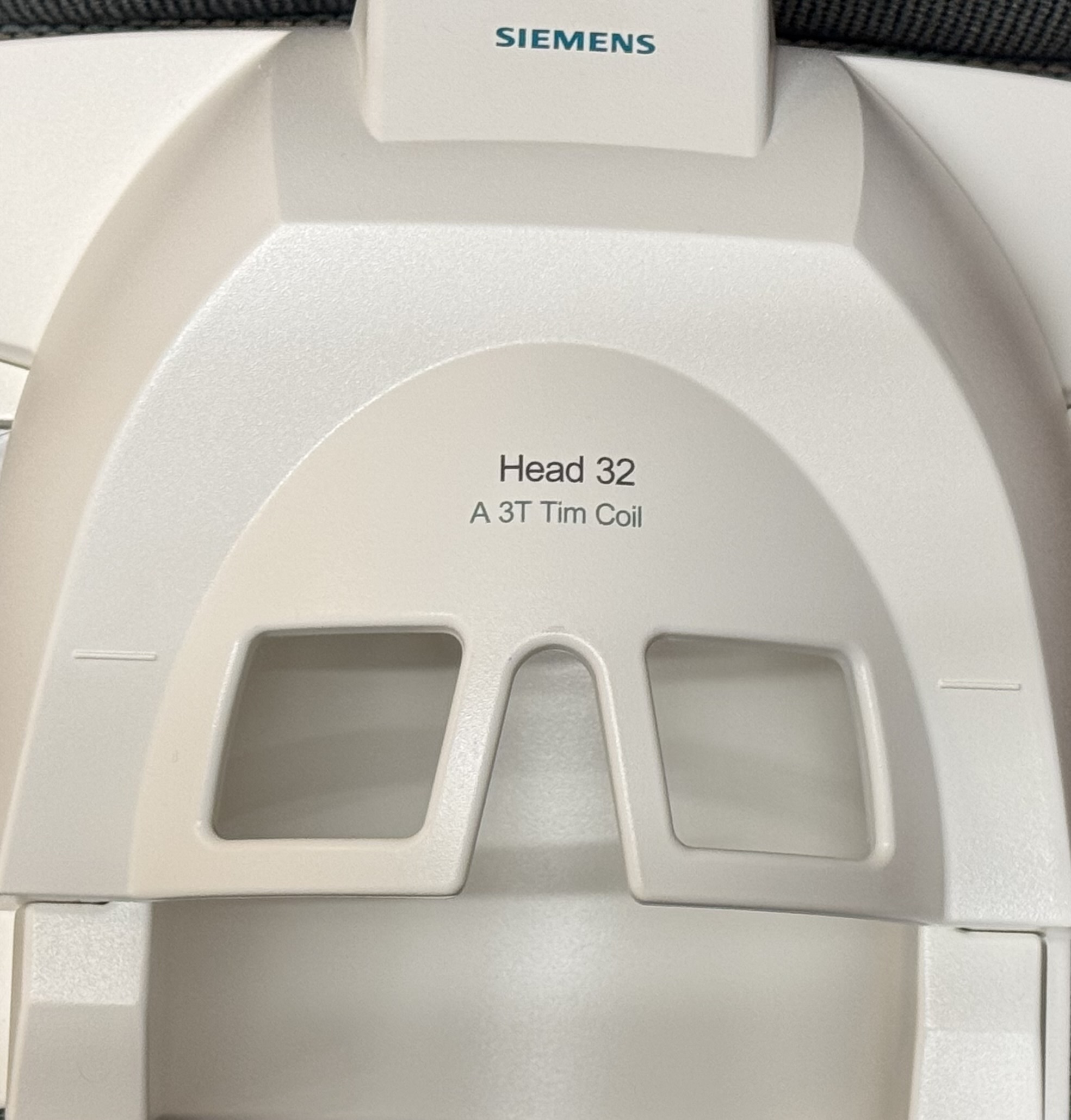}
    \caption{Top view of Head Coil}
    \label{fig:sub_lensholder}
  \end{subfigure}
  \caption{Orthographic views of the components: (a) Front view of Head Coil, (b) Top view of Head Coil. The images of these components were used to understand the geometry of the anterior head coil and help design the head coil mount.}
  \label{fig:grouped_views}
\end{figure}

\begin{figure}[H]
  \centering
  \includegraphics[width=0.4\textwidth]{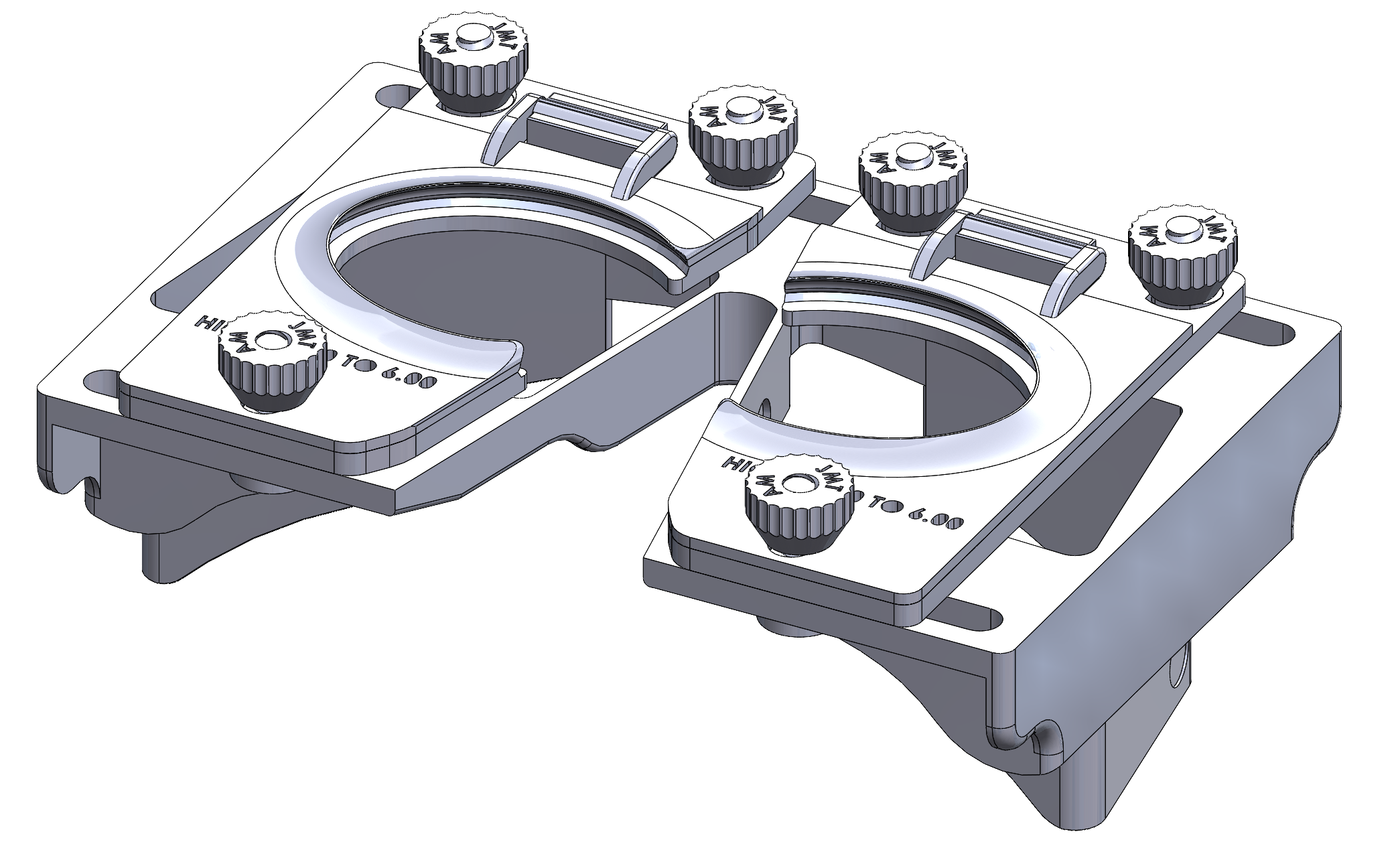}
  \caption{Trimetric view of the Vision Correction Device Assembly. The assembly consists of the head coil mount, lens holders, nuts, and bolts.}
  \label{fig:assembly}
\end{figure}

\begin{figure}[H]
  \centering
  \includegraphics[width=.4\textwidth]{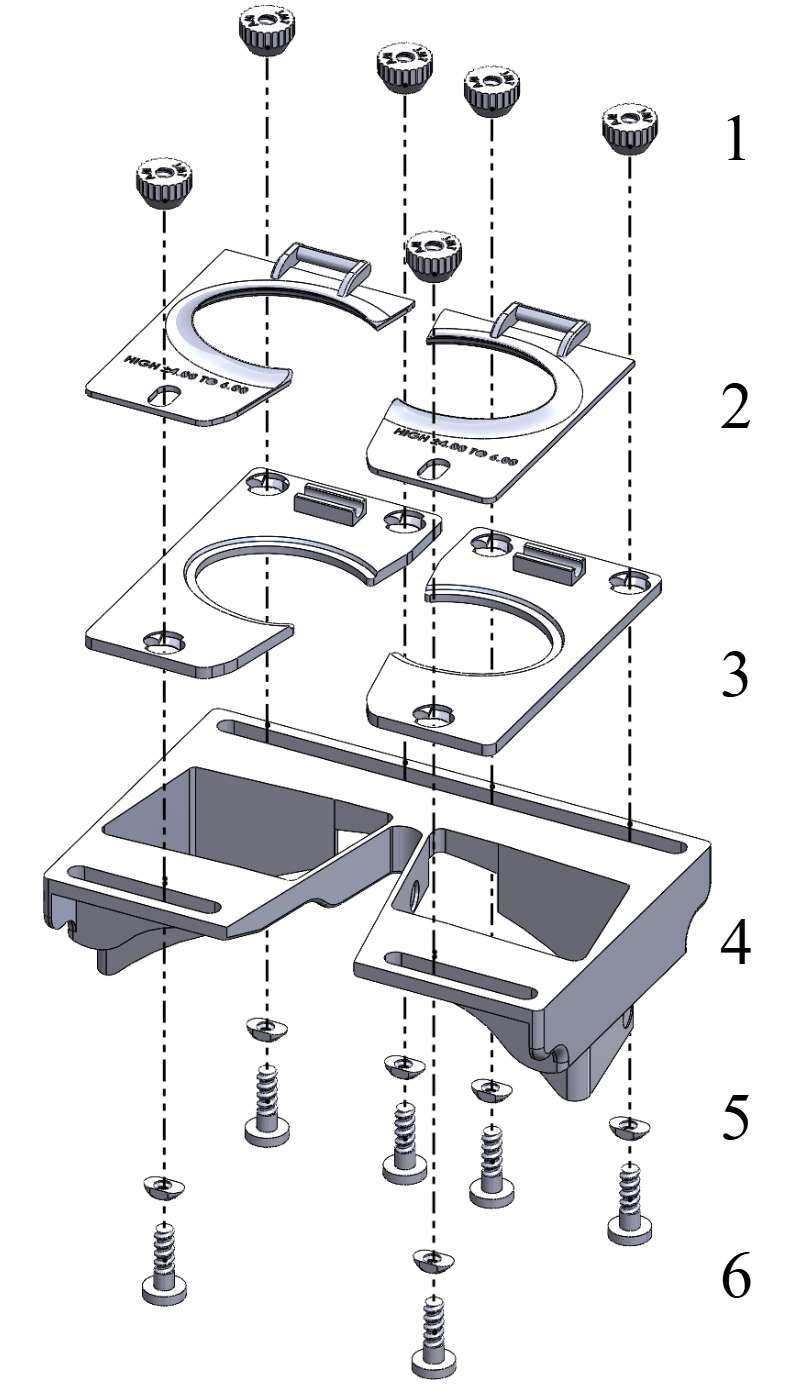}
  \caption{Trimetric view of the Exploded Assembly. This exploded view captures each component of the assembly. 1. Nuts, 2. Lens Holder Top, 3. Lens Holder Bottom, 4. Head Coil Mount, 5. Supporting Washers, 6. Bolts.}
  \label{fig:exploded}
\end{figure}

The device was developed through an iterative prototyping process using SolidWorks Education Edition 2023 and consists of a head coil mount, independent lens holders, and fastening hardware (nuts and bolts), as shown in (Figure~\ref{fig:grouped_parts}). To ensure a precise fit to the Siemens 32-channel head coil, reference photographs were taken from the top and front views and imported into SolidWorks using the Sketch Picture Tool (Figures \ref{fig:Sketch} and \ref{fig:grouped_views}). The complete assembly is illustrated in both assembled (Figure~\ref{fig:assembly}) and exploded views (Figure~\ref{fig:exploded}). Reference physical measurements of the head coil were used to guide accurate sketching directly onto the images. The device was designed to minimize the use of supports to simplify printing and preserve the smooth geometries of the components. Gel dots were added to the design to minimize the effects of acoustic vibrations on the device during scanning protocols.

\subsection{Head Coil Mount}
\begin{figure}[H]
  \centering
  \begin{subfigure}[b]{0.5\textwidth}
    \centering
    \includegraphics[width=\textwidth]{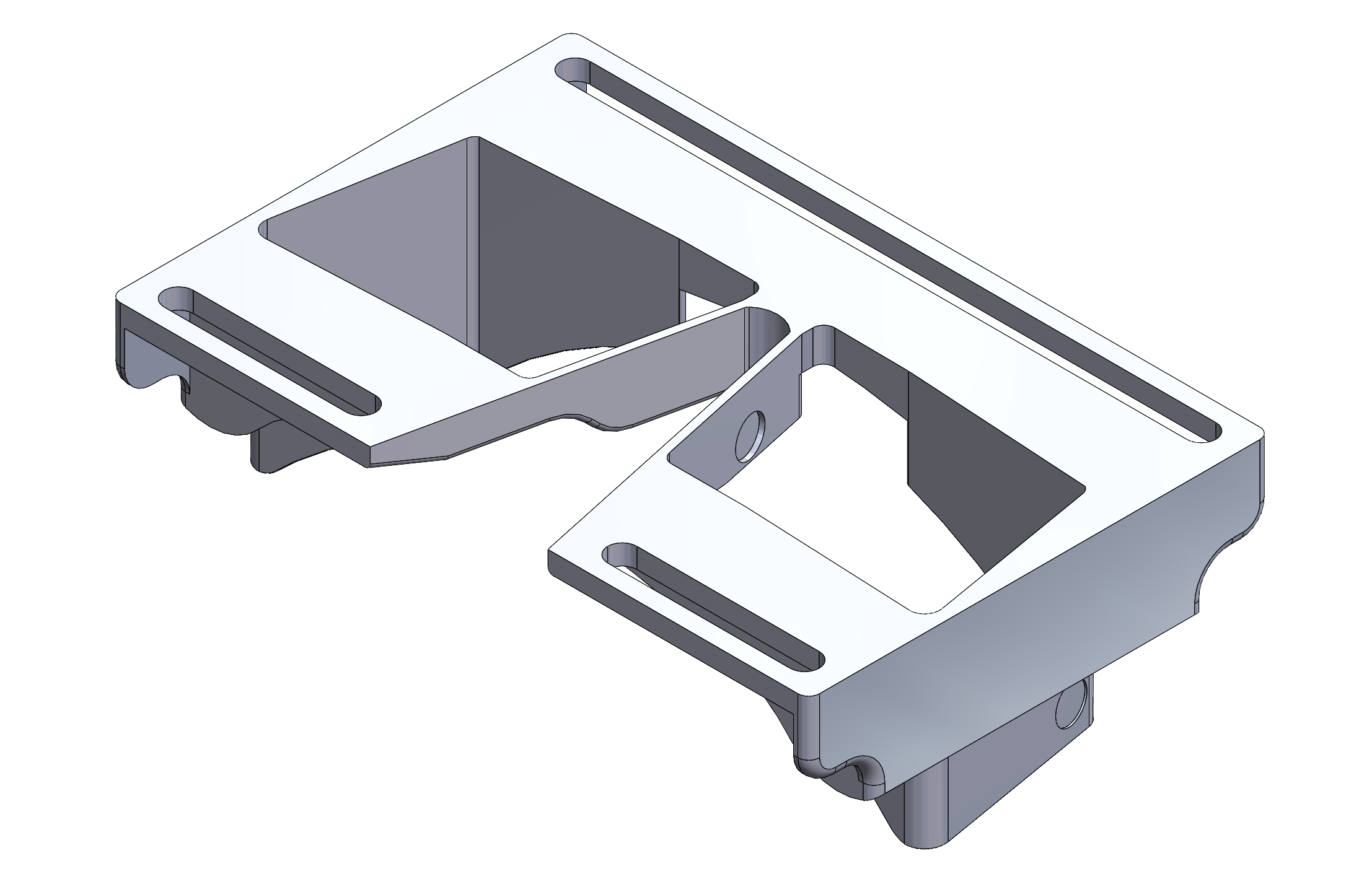}
    \caption{Head Coil Mount}
    \label{fig:sub_headcoilmount_part}
  \end{subfigure}
  \hfill
  \begin{subfigure}[b]{0.4\textwidth}
    \centering
    \includegraphics[width=\textwidth]{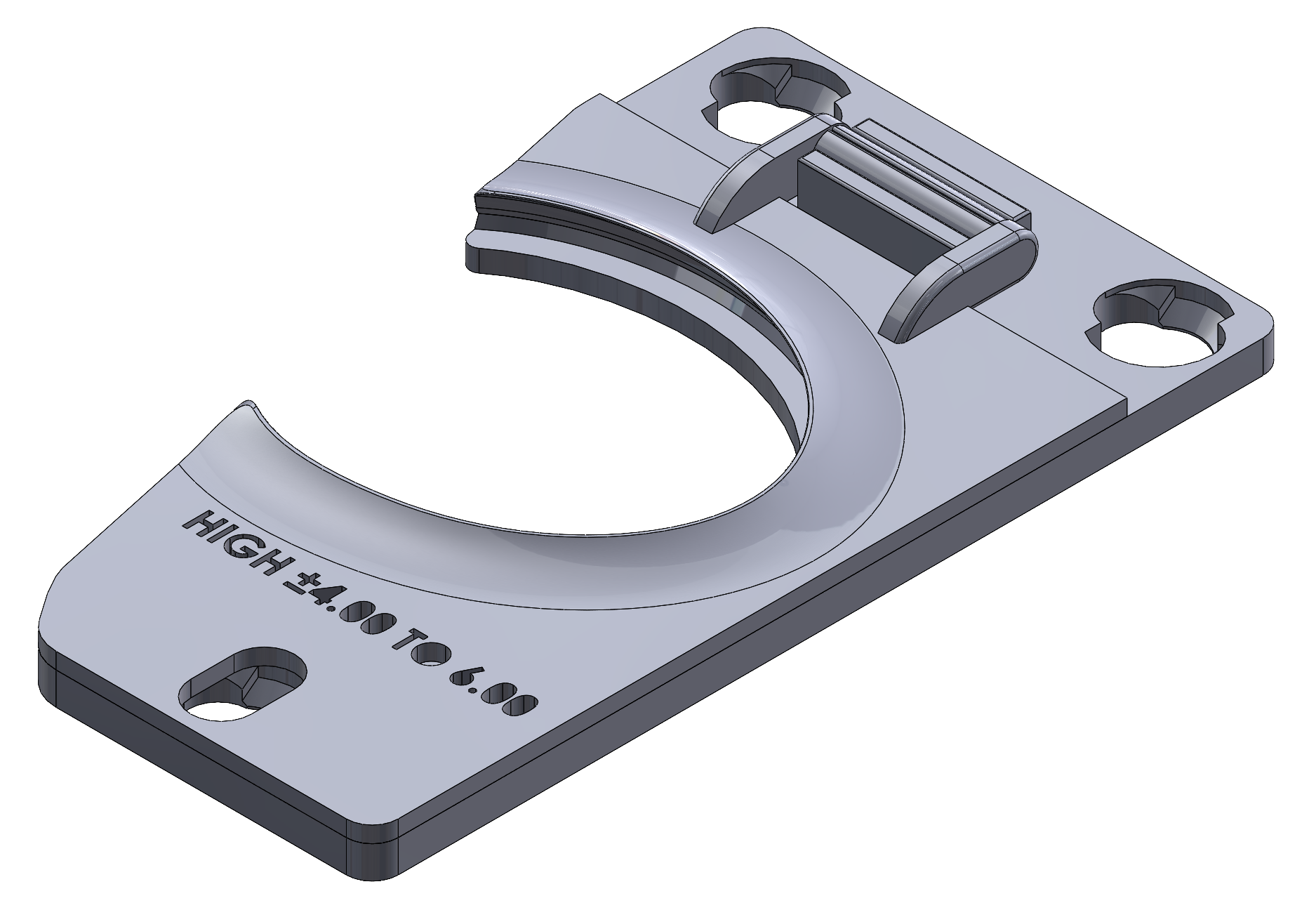}
    \caption{Lens Holder}
    \label{fig:sub_lensholder}
  \end{subfigure}

  \vskip\baselineskip

  \begin{subfigure}[b]{0.2\textwidth}
    \centering
    \includegraphics[width=\textwidth]{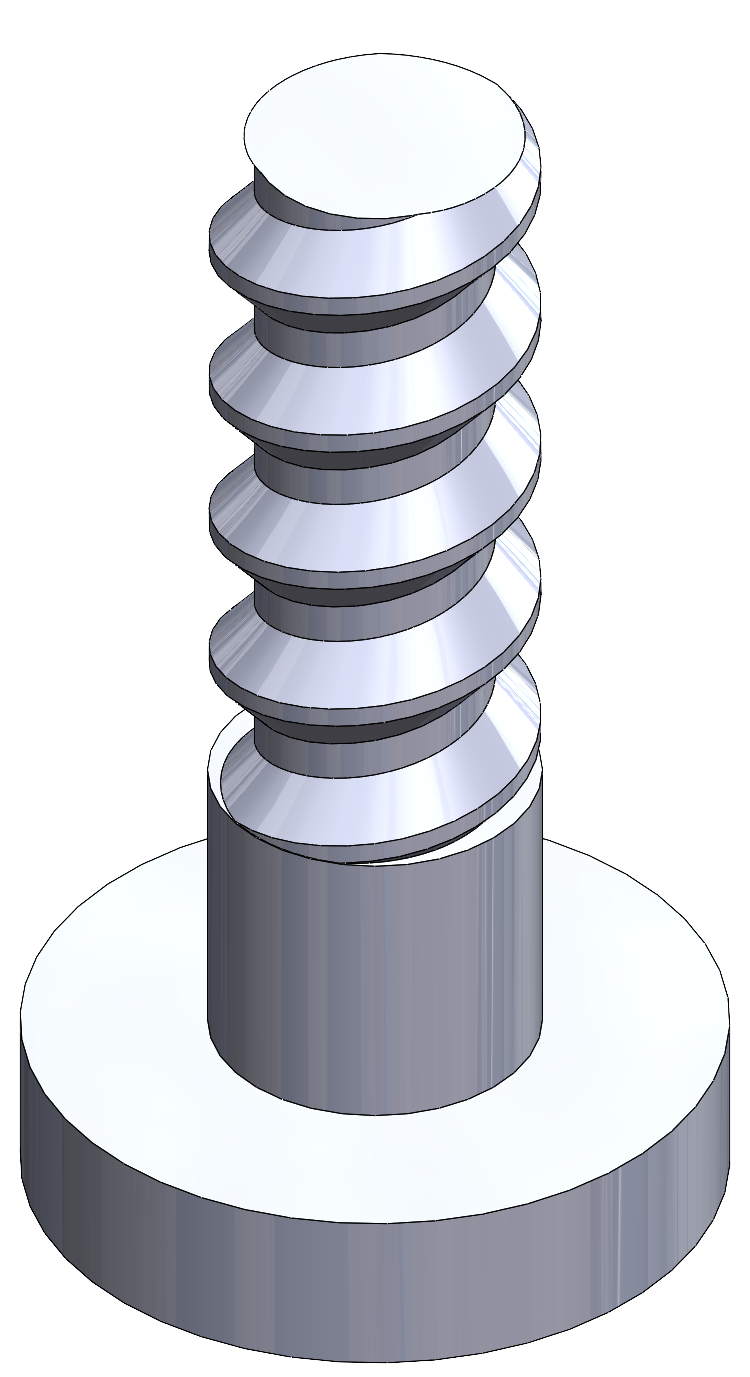}
    \caption{Bolt}
    \label{fig:sub_bolt}
  \end{subfigure}
  \hfill
  \begin{subfigure}[b]{0.2\textwidth}
      \centering
      \includegraphics[width=\linewidth]{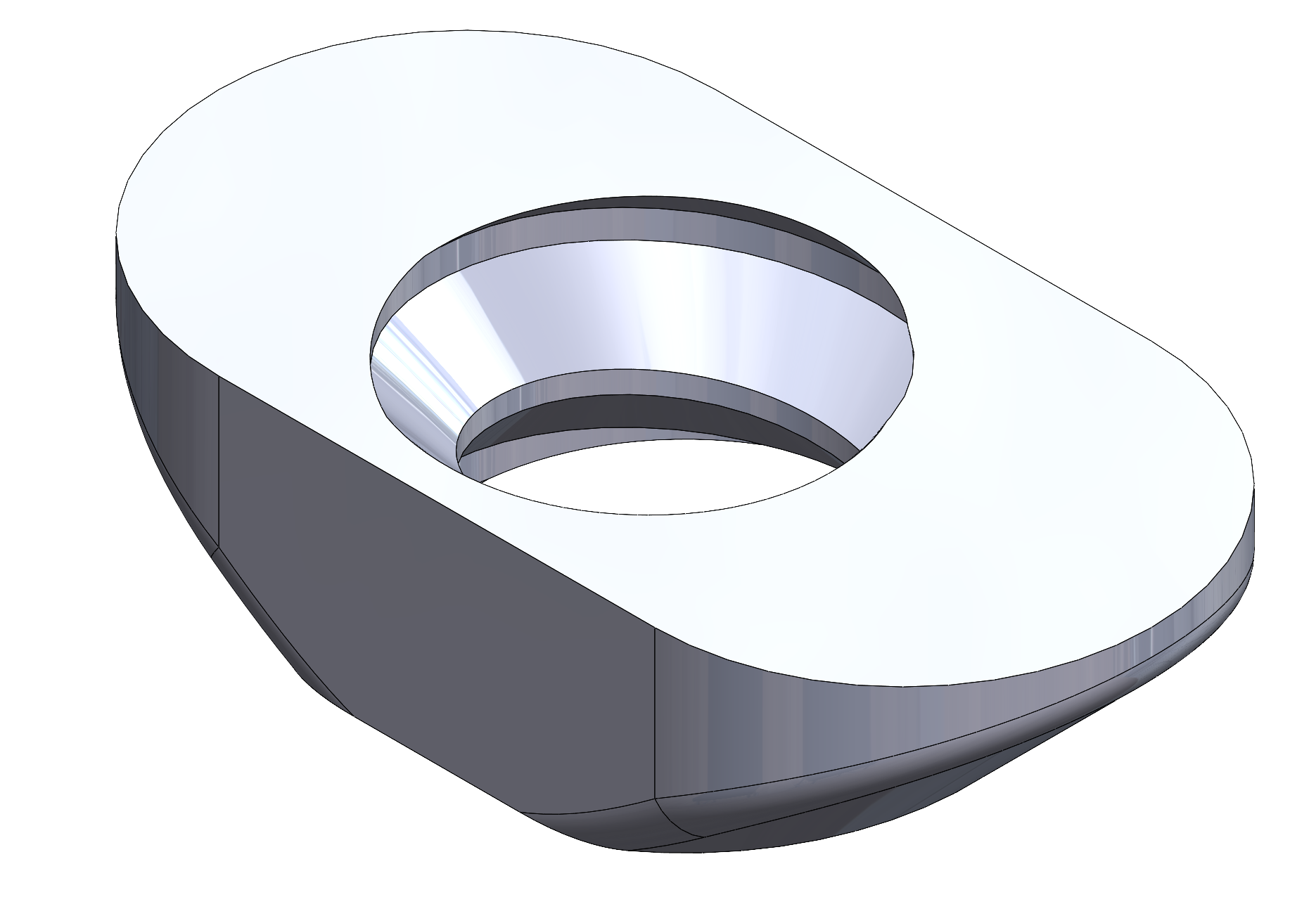}
      \caption{Supporting Washer}
      \label{fig:supporting_washer}
  \end{subfigure}
  \hfill
  \begin{subfigure}[b]{0.2\textwidth}
    \centering
    \includegraphics[width=\textwidth]{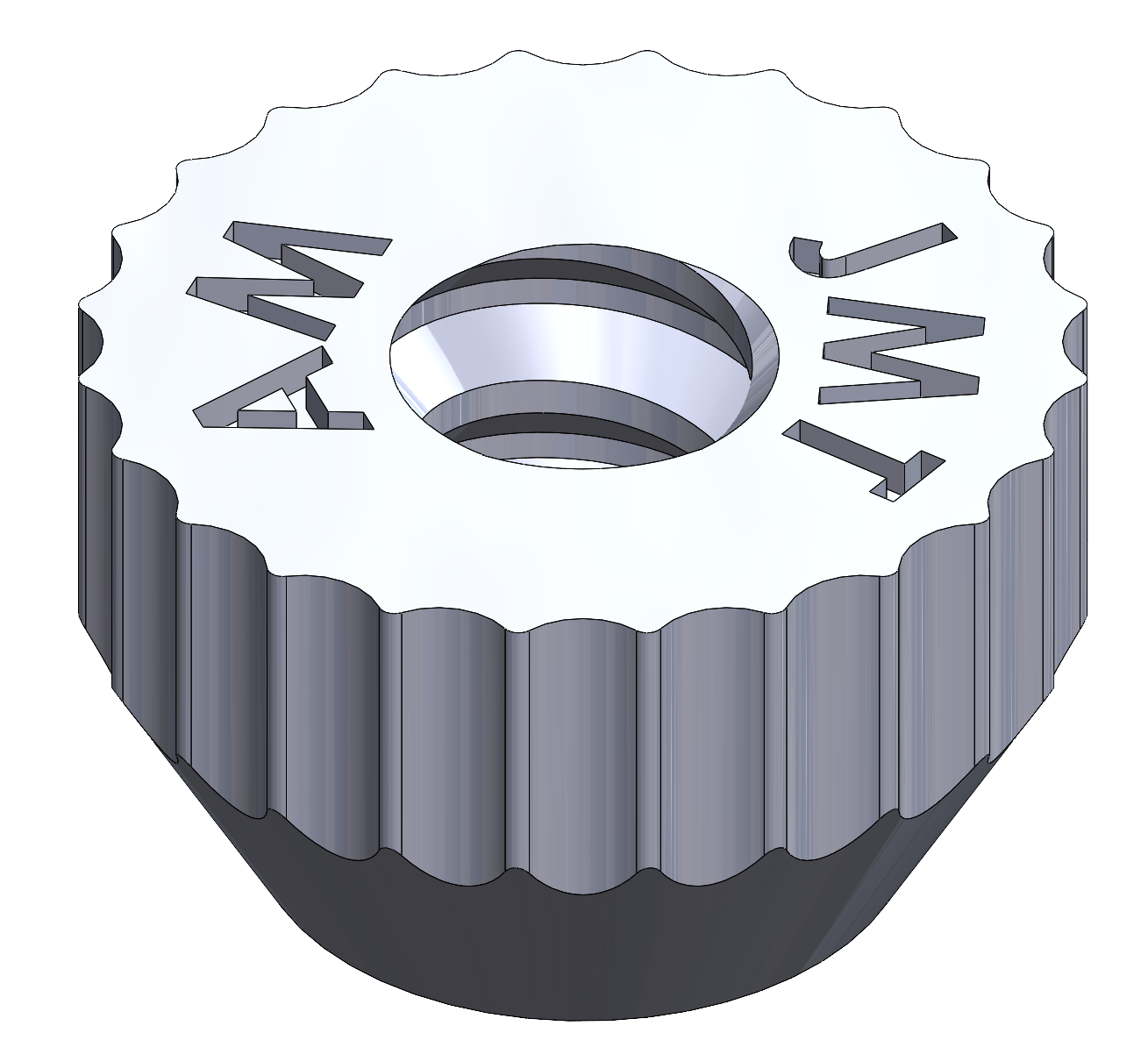}
    \caption{Nut}
    \label{fig:sub_nut}
  \end{subfigure}

  \caption{Isometric views of the components: (a) Head Coil Mount, (b) Lens Holder, (c) Bolt, (d) Supporting Washer, and (e) Nut. Note, components are not to scale.}
  \label{fig:grouped_parts}
\end{figure}

The head coil mount serves as the primary structural component designed to attach securely to the Siemens 32 Channel Head Coil and support the lens holders in a position optimal for the participant's line of sight. The large vertical slots located on the top and bottom of the mount’s front face allow for the stable bolting of the lens holders, as shown in Figure~\ref{fig:sub_headcoilmount_part}. Protruding guide features from the eye slots ensure accurate and stable alignment of the mount within the eye slots of the Siemens 32 Channel Head Coil, as illustrated in Figure~\ref{fig:headcoilmountbottom}.

\subsection{Lens Holder Assembly}
The lens holder assembly secures the corrective lenses using a hinged cover that lifts for lens insertion and removal. It is fastened with three fasteners, with the bottom bolt tightened to lock the cover in place. Three versions of the lens holder assembly were developed to accommodate varying lens thicknesses, with higher diopter corrections typically associated with the thickest lenses.

\subsection{Fasteners}
A total of six fasteners (in this case threaded nuts and bolts, with supporting washers) are required to secure the device. The fasteners can be 3D printed or commercial, non-metallic hardware fabricated from nylon or a similar plastic. The supporting washers are fastened onto the bolts to support the bolts against the lens holder bottoms and fasten the nuts.

\section{Fabrication}
Mesh objects were exported from SolidWorks (Education Edition 2023 SP4.0, Dassault Systèmes, Waltham, MA) and sliced for 3D filament deposition modeling (FDM) using Orca Slicer (version 2.3.0, Orca Slicer, San Antonio, TX). The final device prototype was printed entirely in general purpose polylactic acid (PLA) using an X1 Carbon 3D printer (Bambu Lab, Austin, TX). We recommend using a textured plate for printing all the components described here. Layer heights between 0.08 mm and 0.20 mm are acceptable with 0.08 mm layers providing the highest print quality overall. For improved durability, we used 5 wall loops and 30\% infill density with a cross hatch infill pattern. We recommend using tree supports for the lens holders and mount only and no supports for the nut or bolt. 

\section{Assembly}
\begin{figure}[H]
  \centering
  \includegraphics[width=0.5\textwidth]{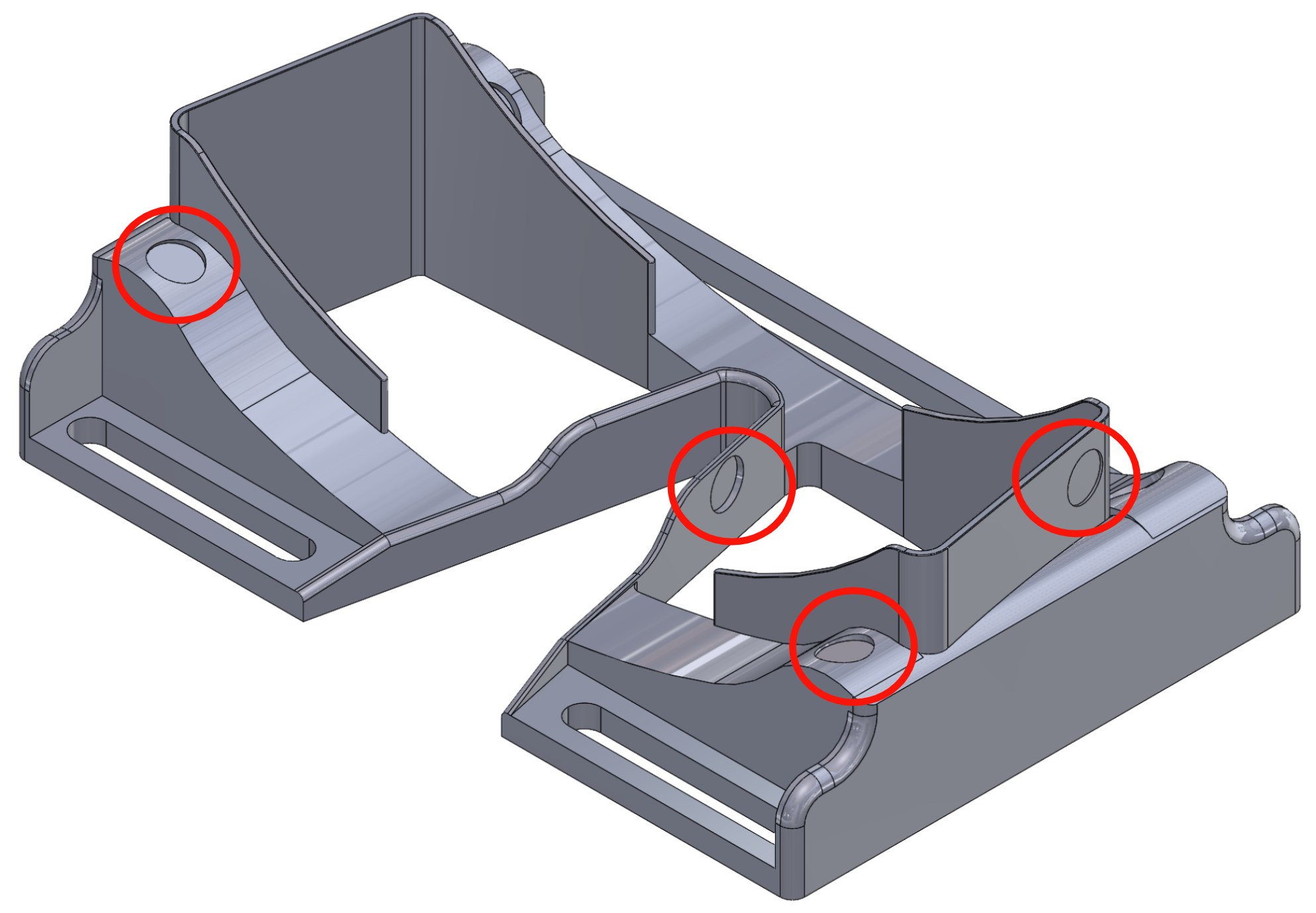} 
  \caption{Isometric view of the bottom of the Head Coil Mount. The red circles indicate the locations of some of the circular recesses to place gel bumpers.}
  \label{fig:headcoilmountbottom}
\end{figure}

\begin{figure}[H]
  \centering
  \includegraphics[width=0.375\textwidth]{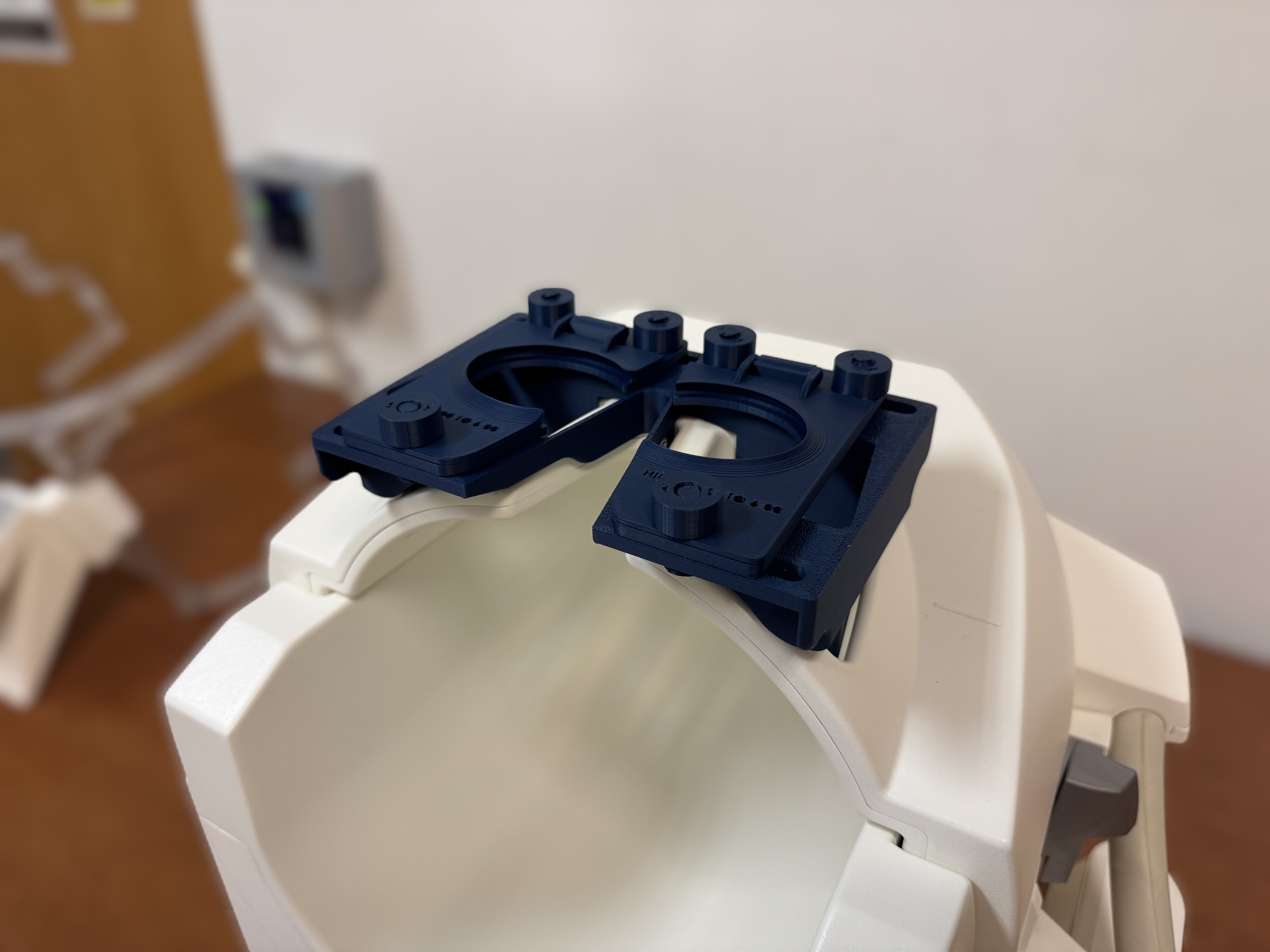} 
  \caption{Isometric view of the Vision Correction Device without lenses mounted on the Head Coil.}
  \label{fig:bareAssembelyOnHeadCoil}
\end{figure}

\begin{figure}[H]
  \centering
  \includegraphics[width=0.375\textwidth]{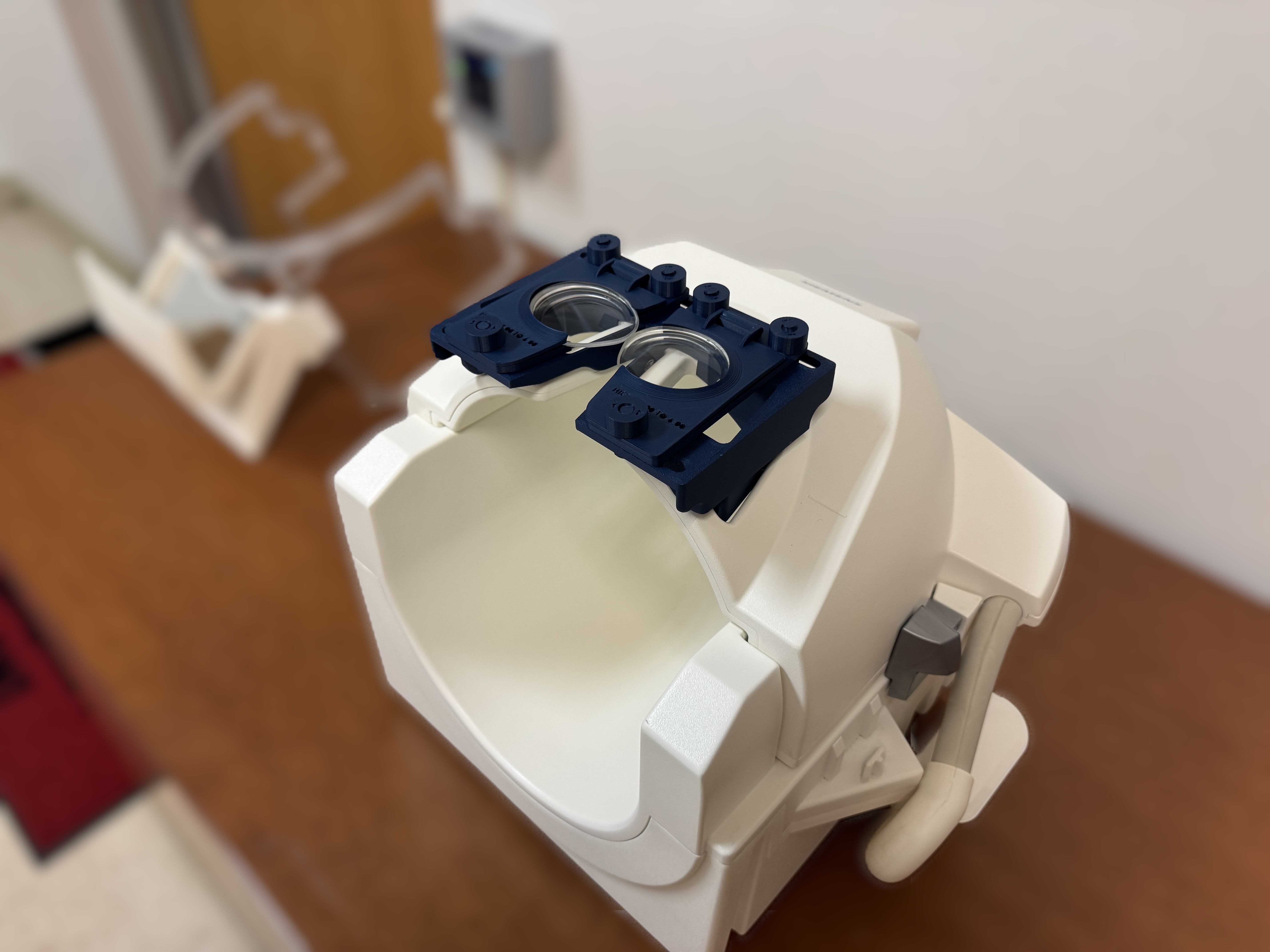} 
  \caption{Isometric view of the Vision Correction Device with lenses mounted on the Head Coil.}
  \label{fig:fullAssembelyOnHeadCoil}
\end{figure}

\begin{figure}[H]
  \centering
  \includegraphics[width=0.375\textwidth]{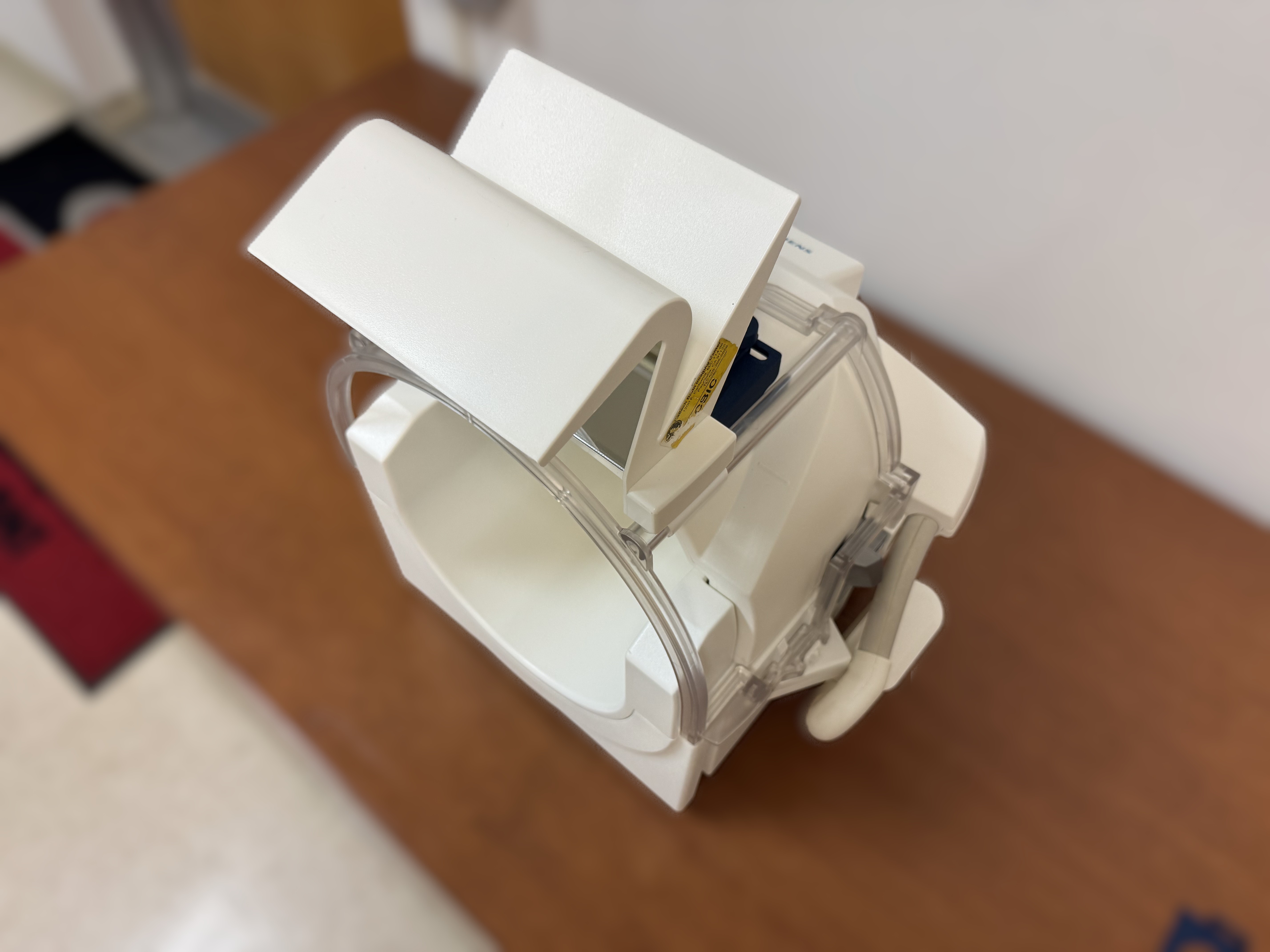} 
  \caption{Isometric view of the Full Vision Correction Device Assembly mounted on the Head Coil.}
  \label{fig:fullAssembelyWithMirrorOnHeadCoil}
\end{figure}

Once parts have been printed, assembly begins with the application of gel bumpers to various circular recesses on lower surface of the head coil mount (Figure \ref{fig:headcoilmountbottom}). We recommend using an appropriate PLA-compatible contact adhesive or gel superglue to attach the bumpers. Once the gel bumpers are secure, place the lens holders onto the mount and fit the bolts through both the head mount slots and the holes in the lens holders. Loosely tighten fasteners to allow for later adjustment. Finally place the assembled vision corrector with lenses in place onto the head coil (Figure \ref{fig:bareAssembelyOnHeadCoil}). We recommend determining the correct lenses for a given participant outside the magnet room using a Snellen chart at 6 meters. Acceptable corrected acuity depends on experimental requirements, but is typically 20/30 to 20/20 (Snellen lines 6 to 8). Place the lenses the participant needs into the lens holders and secure the fasteners loosely (Figure ~\ref{fig:fullAssembelyOnHeadCoil}). Slide the lens holders to the position that provides the best subjective correction for the participant and tighten the remaining fasteners. Finally place the mirror system on the head coil (Figure ~\ref{fig:fullAssembelyWithMirrorOnHeadCoil}). 
    
\subsection{Reference Videos} \label{videos}
We provide a set of reference videos to guide assembly and positioning of the device on the head coil:

\begin{itemize}
    \item \href{https://youtu.be/AsIwOQPe71k}{Gel Placement Youtube Video}
    \item \href{https://youtu.be/XtZz5X_WPjU}{Device Assembly Youtube Video}
   \item \href{https://youtu.be/D7gqrahNJ6g}{Bolt Assembly Youtube Video}
    \item \href{https://youtu.be/qEZhaU0NNYg}{Placing The Lenses}
\end{itemize}

\section{Design Files}
CAD and STL files as well as updated device versions are available at \url{https://github.com/aleks123marinkovic/VisionCorrectionDevice}. If you would like to request modifications or share your updated versions, please contact the corresponding author.

\section{Lens Reflections}
Some lenses may produce reflections that are detectable during eye-tracking. First, clean the lenses with a microfiber cloth to remove oils and smudges that may cause reflections. If the reflections persist, it may be possible to find a professional to coat your lenses with an anti-reflective coating, but it is ideal to use lenses that come with an anti-reflective coating. Since the design of the lens holders are meant to be modular, it is easy to design lens holders for the shape of lenses with an anti-reflective coating.

\section{Other Use Cases}
This device can serve a multitude of purposes besides vision correction. It can also be used in vision experiments by occluding or partially inhibiting vision in one or both eyes by developing alternatives to the lens holders to sit atop the mounting face.

\section{Acknowledgments}
The authors would like to thank Arleen Hom and Ian Roberts (Caltech Library) for their assistance with 3D printing and material selection. We are also grateful to Professor Paul Stovall (Division of Mechanical and Civil Engineering, Caltech) for his feedback on the mechanical drawings. 

\printbibliography
\newpage
\begin{table}[H]
\small
\renewcommand{\arraystretch}{1.2}
\centering
\begin{tabularx}{\textwidth}{@{} >{\raggedright\arraybackslash}p{0.28\textwidth}
                                >{\raggedright\arraybackslash}p{0.18\textwidth}
                                >{\raggedright\arraybackslash}p{0.2\textwidth}
                                L @{}}
\toprule
\textbf{Design file name} & \textbf{File type} & \textbf{License} & \textbf{Location of the file} \\
\midrule

Lens\_Holder\_Left\_Bottom.SLDPRT & SLDPRT & CC BY-NC-SA 4.0 &
\href{https://github.com/aleks123marinkovic/VisionCorrectionDevice/blob/main/Part_Files/Lens_Holder_Parts/Lens_Holder_Left_Bottom.SLDPRT}{github link} \\

Lens\_Holder\_Right\_Bottom.SLDPRT & SLDPRT & CC BY-NC-SA 4.0 &
\href{https://github.com/aleks123marinkovic/VisionCorrectionDevice/blob/main/Part_Files/Lens_Holder_Parts/Lens_Holder_Right_Bottom.SLDPRT}{github link} \\

Lens\_Holder\_Left\_High.SLDPRT & SLDPRT & CC BY-NC-SA 4.0 &
\href{https://github.com/aleks123marinkovic/VisionCorrectionDevice/blob/main/Part_Files/Lens_Holder_Parts/Lens_Holder_Left_High.SLDPRT}{github link} \\

Lens\_Holder\_Right\_High.SLDPRT & SLDPRT & CC BY-NC-SA 4.0 &
\href{https://github.com/aleks123marinkovic/VisionCorrectionDevice/blob/main/Part_Files/Lens_Holder_Parts/Lens_Holder_Right_High.SLDPRT}{github link} \\

Lens\_Holder\_Left\_Med.SLDPRT & SLDPRT & CC BY-NC-SA 4.0 &
\href{https://github.com/aleks123marinkovic/VisionCorrectionDevice/blob/main/Part_Files/Lens_Holder_Parts/Lens_Holder_Left_Med.SLDPRT}{github link} \\

Lens\_Holder\_Right\_Med.SLDPRT & SLDPRT & CC BY-NC-SA 4.0 &
\href{https://github.com/aleks123marinkovic/VisionCorrectionDevice/blob/main/Part_Files/Lens_Holder_Parts/Lens_Holder_Right_Med.SLDPRT}{github link} \\

Lens\_Holder\_Left\_Low.SLDPRT & SLDPRT & CC BY-NC-SA 4.0 &
\href{https://github.com/aleks123marinkovic/VisionCorrectionDevice/blob/main/Part_Files/Lens_Holder_Parts/Lens_Holder_Left_Low.SLDPRT}{github link} \\

Lens\_Holder\_Right\_Low.SLDPRT & SLDPRT & CC BY-NC-SA 4.0 &
\href{https://github.com/aleks123marinkovic/VisionCorrectionDevice/blob/main/Part_Files/Lens_Holder_Parts/Lens_Holder_Right_Low.SLDPRT}{github link} \\

Mount.SLDPRT & SLDRT & CC BY-NC-SA 4.0 &
\href{https://github.com/aleks123marinkovic/VisionCorrectionDevice/tree/main/Part_Files/Mount_Parts}{github link} \\

Bolt.SLDPRT & SLDRT & CC BY-NC-SA 4.0 &
\href{https://github.com/aleks123marinkovic/VisionCorrectionDevice/blob/main/Part_Files/Nut_And_Bolt_And_Washer_Parts/Bolt.SLDPRT}{github link} \\

Nut.SLDPRT & SLDRT & CC BY-NC-SA 4.0 &
\href{https://github.com/aleks123marinkovic/VisionCorrectionDevice/blob/main/Part_Files/Nut_And_Bolt_And_Washer_Parts/Nut.SLDPRT}{github link} \\

Supporting\_Washer.SLDPRT & SLDRT & CC BY-NC-SA 4.0 &
\href{https://github.com/aleks123marinkovic/VisionCorrectionDevice/blob/main/Part_Files/Nut_And_Bolt_And_Washer_Parts/Supporting_Washer.SLDPRT}{github link} \\

Lens\_Holder\_Bottoms.3mf & 3mf & CC BY-NC-SA 4.0 &
\href{https://github.com/aleks123marinkovic/VisionCorrectionDevice/blob/main/Print_Files/3mf_Files/Lens_Holder_Bottoms.3mf}{github link} \\

Lens\_Holders\_Left.3mf & 3mf & CC BY-NC-SA 4.0 &
\href{https://github.com/aleks123marinkovic/VisionCorrectionDevice/blob/main/Print_Files/3mf_Files/Lens_Holders_Left.3mf}{github link} \\

Lens\_Holders\_Right.3mf & 3mf & CC BY-NC-SA 4.0 &
\href{https://github.com/aleks123marinkovic/VisionCorrectionDevice/blob/main/Print_Files/3mf_Files/Lens_Holders_Right.3mf}{github link} \\

Mount.3mf & 3mf & CC BY-NC-SA 4.0 &
\href{https://github.com/aleks123marinkovic/VisionCorrectionDevice/blob/main/Print_Files/3mf_Files/Mount.3mf}{github link} \\

Bolts.3mf & 3mf & CC BY-NC-SA 4.0 &
\href{https://github.com/aleks123marinkovic/VisionCorrectionDevice/blob/main/Print_Files/3mf_Files/Bolts.3mf}{github link} \\

Nuts.3mf & 3mf & CC BY-NC-SA 4.0 &
\href{https://github.com/aleks123marinkovic/VisionCorrectionDevice/blob/main/Print_Files/3mf_Files/Nuts.3mf}{github link} \\

Supporting\_Washer.3mf & 3mf & CC BY-NC-SA 4.0 &
\href{https://github.com/aleks123marinkovic/VisionCorrectionDevice/blob/main/Print_Files/3mf_Files/Supporting_Washer.3mf}{github link} \\

Lens\_Holder\_Left\_Bottom.STEP & STEP & CC BY-NC-SA 4.0 &
\href{https://github.com/aleks123marinkovic/VisionCorrectionDevice/blob/main/Print_Files/STEP_Files/Lens_Holder_Left_Bottom.STEP}{github link} \\

Lens\_Holder\_Right\_Bottom.STEP & STEP & CC BY-NC-SA 4.0 &
\href{https://github.com/aleks123marinkovic/VisionCorrectionDevice/blob/main/Print_Files/STEP_Files/Lens_Holder_Right_Bottom.STEP}{github link} \\

Lens\_Holder\_Left\_High.STEP & STEP & CC BY-NC-SA 4.0 &
\href{https://github.com/aleks123marinkovic/VisionCorrectionDevice/blob/main/Print_Files/STEP_Files/Lens_Holder_Left_High.STEP}{github link} \\

Lens\_Holder\_Right\_High.STEP & STEP & CC BY-NC-SA 4.0 &
\href{https://github.com/aleks123marinkovic/VisionCorrectionDevice/blob/main/Print_Files/STEP_Files/Lens_Holder_Right_High.STEP}{github link} \\

Lens\_Holder\_Left\_Med.STEP & STEP & CC BY-NC-SA 4.0 &
\href{https://github.com/aleks123marinkovic/VisionCorrectionDevice/blob/main/Print_Files/STEP_Files/Lens_Holder_Left_Med.STEP}{github link} \\

Lens\_Holder\_Right\_Med.STEP & STEP & CC BY-NC-SA 4.0 &
\href{https://github.com/aleks123marinkovic/VisionCorrectionDevice/blob/main/Print_Files/STEP_Files/Lens_Holder_Right_Med.STEP}{github link} \\

Lens\_Holder\_Left\_Low.STEP & STEP & CC BY-NC-SA 4.0 &
\href{https://github.com/aleks123marinkovic/VisionCorrectionDevice/blob/main/Print_Files/STEP_Files/Lens_Holder_Left_Low.STEP}{github link} \\

Lens\_Holder\_Right\_Low.STEP & STEP & CC BY-NC-SA 4.0 &
\href{https://github.com/aleks123marinkovic/VisionCorrectionDevice/blob/main/Print_Files/STEP_Files/Lens_Holder_Right_Low.STEP}{github link} \\

Mount.STEP & STEP & CC BY-NC-SA 4.0 &
\href{https://github.com/aleks123marinkovic/VisionCorrectionDevice/blob/main/Print_Files/STEP_Files/Mount.STEP}{github link} \\

Bolt.STEP & STEP & CC BY-NC-SA 4.0 &
\href{https://github.com/aleks123marinkovic/VisionCorrectionDevice/blob/main/Print_Files/STEP_Files/Bolt.STEP}{github link} \\

Nut.STEP & STEP & CC BY-NC-SA 4.0 &
\href{https://github.com/aleks123marinkovic/VisionCorrectionDevice/blob/main/Print_Files/STEP_Files/Nut.STEP}{github link} \\

Supporting\_Washer.STEP & STEP & CC BY-NC-SA 4.0 &
\href{https://github.com/aleks123marinkovic/VisionCorrectionDevice/blob/main/Print_Files/STEP_Files/Supporting_Washer.STEP}{github link} \\

Lens\_Holder\_Left\_Bottom.STL & STL & CC BY-NC-SA 4.0 &
\href{https://github.com/aleks123marinkovic/VisionCorrectionDevice/blob/main/Print_Files/STL_Files/Lens_Holder_Left_Bottom.STL}{github link} \\

Lens\_Holder\_Right\_Bottom.STL & STL & CC BY-NC-SA 4.0 &
\href{https://github.com/aleks123marinkovic/VisionCorrectionDevice/blob/main/Print_Files/STL_Files/Lens_Holder_Right_Bottom.STL}{github link} \\

Lens\_Holder\_Left\_High.STL & STL & CC BY-NC-SA 4.0 &
\href{https://github.com/aleks123marinkovic/VisionCorrectionDevice/blob/main/Print_Files/STL_Files/Lens_Holder_Left_High.STL}{github link} \\

Lens\_Holder\_Right\_High.STL & STL & CC BY-NC-SA 4.0 &
\href{https://github.com/aleks123marinkovic/VisionCorrectionDevice/blob/main/Print_Files/STL_Files/Lens_Holder_Right_High.STL}{github link} \\

Lens\_Holder\_Left\_Med.STL & STL & CC BY-NC-SA 4.0 &
\href{https://github.com/aleks123marinkovic/VisionCorrectionDevice/blob/main/Print_Files/STL_Files/Lens_Holder_Left_Med.STL}{github link} \\

Lens\_Holder\_Right\_Med.STL & STL & CC BY-NC-SA 4.0 &
\href{https://github.com/aleks123marinkovic/VisionCorrectionDevice/blob/main/Print_Files/STL_Files/Lens_Holder_Right_Med.STL}{github link} \\

Lens\_Holder\_Left\_Low.STL & STL & CC BY-NC-SA 4.0 &
\href{https://github.com/aleks123marinkovic/VisionCorrectionDevice/blob/main/Print_Files/STL_Files/Lens_Holder_Left_Low.STL}{github link} \\

Lens\_Holder\_Left\_Low.STL & STL & CC BY-NC-SA 4.0 &
\href{https://github.com/aleks123marinkovic/VisionCorrectionDevice/blob/main/Print_Files/STL_Files/Lens_Holder_Right_Low.STL}{github link} \\

Mount.STL & STL & CC BY-NC-SA 4.0 &
\href{https://github.com/aleks123marinkovic/VisionCorrectionDevice/blob/main/Print_Files/STL_Files/Mount.STL}{github link} \\

Bolt.STL & STL & CC BY-NC-SA 4.0 &
\href{https://github.com/aleks123marinkovic/VisionCorrectionDevice/blob/main/Print_Files/STL_Files/Bolt.STL}{github link} \\

Nut.STL & STL & CC BY-NC-SA 4.0 &
\href{https://github.com/aleks123marinkovic/VisionCorrectionDevice/blob/main/Print_Files/STL_Files/Nut.STL}{github link} \\

Supporting\_Washer.STL & STL & CC BY-NC-SA 4.0 &
\href{https://github.com/aleks123marinkovic/VisionCorrectionDevice/blob/main/Print_Files/STL_Files/Supporting_Washer.STL}{github link} \\

\bottomrule
\end{tabularx}
\normalsize
\end{table}

\end{document}